\documentclass[aps,prb,twocolumn,groupedaddress,amsmath,amsfonts,showpacs]{revtex4}

\usepackage{graphicx}
\usepackage{bm} 
\usepackage{braket}
\usepackage[usenames]{color}

\begin{document}


\title{ 
Green's Function Method for Line Defects and Gapless Modes in Topological Insulators
: Beyond Semi-classical Approach}


\author{Ken Shiozaki}
\author{Satoshi Fujimoto}
\affiliation{Department of Physics, Kyoto University, Kyoto 606-8502, Japan}




\date{\today}

\begin{abstract}
Defects which appear in heterostructure junctions involving topological insulators are
sources of gapless modes governing the low energy properties  of the systems,
as recently elucidated by Teo and Kane [Physical Review B82, 115120 (2010)].
A standard approach for the calculation of topological invariants associated with defects is
to deal with the spatial inhomogeneity raised by defects within a semiclassical approximation. 
In this paper, we propose a full quantum formulation for the topological invariants 
characterizing line defects in three-dimensional insulators with no symmetry by using 
the Green's function method. 
On the basis of the full quantum treatment,
we demonstrate the existence of a nontrivial topological invariant in 
the topological insulator-ferromagnet tri-junction systems, 
for which a semiclassical approximation fails to describe the topological phase.
Also, our approach enables us to study effects of electron-electron interactions and impurity scattering on topological insulators with spatial
inhomogeneity which gives rise to the Axion electrodynamics responses.

\end{abstract}



\maketitle



\section{Introduction}
\label{sec1}

Topological phases realized in condensed matter systems have been currently attracting much interest.
\cite{TKNN,Volovik,HasanKane,QiZhang,KM,KM2,FK,Murakami,FKM,FK2,moore,BHZ,Konig,HQW,QHZ,HX,ZLQ,CHEN,XWQ,sheng,Fukui1,Fukui2,FK3,Schnyder,Kitaev,ryu,Chung,Roy,Sato1,SF,STF,Sato2,Sau,Alicea,ANB,Tanaka,Nomura,Santos}
The topological classification of the ground states of bulk band insulators or superconductors 
is a useful approach for  the elucidation of 
low energy properties of the topological phases.
The fundamental consequence of this classification is 
the {\it bulk-boundary correspondence}; i.e. 
there exist the topologically protected gapless states on a
boundary surface which separates two systems belonging to different topological classes.\cite{Schnyder,Kitaev}    
Recently, Teo and Kane proposed the generalization of this topological classification to 
arbitrary topological defects such as line defects and point defects in insulators 
and superconductors.\cite{TK} 
It was suggested that the existence of gapless modes in defects follows from
topologically nontrivial ground states of 
Hamiltonians which vary with material parameters characterizing the defects. 
The Teo-Kane theory was successfully applied to various kinds of heterostructure junction systems, 
describing correctly low energy gapless modes which appear at junctions.
A basic idea of Teo-Kane theory is an "adiabatic deformation". 
The interface structure with a finite energy gap between the ground state and the first excited state 
can be deformed smoothly into the Hamiltonian varies slowly in real space 
without closing the energy gap. 
Then, dealing with the spatial coordinates $\bm{r}$ which parametrize defects as
adiabatic parameters, one can introduce the adiabatic Hamiltonian $H(\bm{k},\bm{r})$,
which can be used for the argument on topological properties of defects.
Although the underlying idea is quite general, and independent of approximation schemes of calculations,
a simple method for implementing this idea is to apply 
the semi-classical approximation to such a slowly varying Hamiltonian, 
and to treat the spatial coordinate $\bm{r}$  
and the momentum $\bm{k}$ as independent variables.
In fact, this approach was adopted for the argument on the topological insulator-antiferromagnet junction 
in ref.\cite{TK}.
%
The semi-classical approach is particularly useful when the spatial variation of material 
parameters raised by defects is sufficiently slow.
In the case of line defects with broken time reversal symmetry, which
may be raised by the topological insulator-magnet junctions,
the topological invariant within a semi-classical approximation 
can be represented by the winding number of the axion
$\theta$-term \cite{QHZ} which characterizes the topological structures of band insulators in bulk regimes. 
As long as the spatial variation of material parameters in the vicinity of defects is sufficiently slow,
we can calculate the $\theta$-term
by using the semi-classical approach.
\cite{TK,QHZ,EMV} 

However, in some cases, a semi-classical approximation 
fails to give
the correct energy spectrum of spatially inhomogeneous systems.
For instance, in the case of a topological insulator-ferromagnet heterostructure junction,
which is particularly important for the application to the quantum Hall effect,
a semi-classical approximation which neglects quantum corrections leads to the closing of
an energy gap at the interface between the topological insulator and the ferromagnet.
Since the full quantum energy spectrum of this system has an energy gap,
the gap-closing is a spurious effect raised by the semiclassical approximation.
Because of the gap-closing, we can not evaluate
the topological invariant, and hence, the semiclassical approach for
the topological classification of defects is flawed.

Motivated by the above consideration,
we propose a full quantum formulation for the topological invariant 
characterizing line defects in three-dimensional insulators with no symmetry exploiting
the Green's function method, 
but, without using both adiabatic deformation and semi-classical approximation.
On the basis of the full quantum treatment,
we demonstrate the existence of a nontrivial topological invariant in 
the topological insulator-ferromagnet tri-junction systems of which the energy gap is closed within 
the semi-classical approximation.

The organization of this paper is as follows: 
in Sec.\ref{sec2}, we briefly review the Teo-Kane classification scheme, particularly focusing on the topological 
invariant for line defects in a three-dimensional insulator. 
We also demonstrate that in a topological insulator-ferromagnet heterostructure system a semi-classical 
approximation neglecting quantum corrections to the energy spectrum fails to
capture the correct topological features of the system. 
In Sec.\ref{sec3}, we present the full quantum construction of the topological invariant for line defects
with  broken time-reversal symmetry. 
In Sec.\ref{sec4} and Sec.\ref{sec5}, we apply our approach to
the topological insulator-ferromagnet heterostructure system. 
For this purpose, we use an exactly solvable model of the heterostructure.
We present the exact solution of the heterostructure system in Sec.\ref{sec4}, and then,
show the numerical results of the topological invariants
in Sec.\ref{sec5}.
In Sec.\ref{sec6}, we give a conclusion and discussions.

\section{The Teo-Kane's theory and The Semi-classical Approach}
\label{sec2}

The Teo-Kane theory assumes the semi-classical Hamiltonians $H(\bm{k},\bm{r})$ which vary 
slowly in real space far away from defects. 
The existence of gapless modes localized at the defects follows from
topologically nontrivial ground states of Hamiltonians 
$H(\bm{k},\bm{s}) := H \left(\bm{k},\bm{r}(\bm{s})\right)$,
where $\bm{s}$ parametrizes a line or a surface
surrounding the defects. 
A similar semi-classical approach to the classification of topological defects was also 
considered for the superfluid ${}^3$He.\cite{Volovik}  
In the cases of line defects in three dimensional insulators with no symmetry, 
the classification of the ground states of the semi-classical Hamiltonians $H(k_x,k_y,k_z,s)$ is 
an integer $\mathbb{Z}$ characterized by the second Chern number 
\begin{equation}
\begin{split}
Ch_2 = - \frac{1}{8 \pi^2} \int_{T^3 \times S^1} \mathrm{tr}\left[ \mathcal{F} \wedge \mathcal{F} \right], 
\end{split}
\label{second_Chern_number}
\end{equation}
here $\mathcal{F} = d \mathcal{A} + \mathcal{A} \wedge \mathcal{A}$ is the Berry curvature form
\cite{Nakahara}
associated with the non-Abelian Berry connection  
$\mathcal{A}_{i j } = \braket{u_i | d u_j}$ characterizing the valence band 
eigenstates $\ket{u_i (\bm{k},s)}$. 
The second Chern number $Ch_2$ can be represented as the winding number of the $\theta$-term 
describing the local band structure 
\begin{equation}
\begin{split}
Ch_2 &= \frac{1}{2 \pi} \oint_{S^1} ds \frac{d}{ds} \theta (s),
\end{split}
\end{equation}
\begin{equation}
\label{theta_semi-classical}
\begin{split}
\theta (s) &= 2 \pi \int d^3 \bm{k} \ \mathcal{Q}_3 (\bm{k},s),
\end{split}
\end{equation}
where $\mathcal{Q}_3 (\bm{k},s)$ is the Chern-Simons 3 form 
$\mathcal{Q}_3 = -\frac{1}{8 \pi^2} \mathrm{tr}
\left[ \mathcal{A} \wedge d \mathcal{A} + \frac{2}{3} \mathcal{A} \wedge \mathcal{A} 
\wedge \mathcal{A} \right]$. 
Then the nonzero winding number of $\theta$ corresponds to the existence of gapless modes. 
The $\theta$-term is the same as the axion field in the context of the axion electrodynamics 
describing the electromagnetic property of the topological insulators.\cite{QHZ} 

The above argument is based on a semi-classical approach because quantum corrections 
to the energy spectrum
from 
spatial inhomogeneities are neglected. 
Hence 
it is not guaranteed
that eq.(1) can describe the existence of gapless modes 
in cases where the semi-classical approximation fails.
In fact, this may be the case when
the scale of the spatial variation in the vicinity of interfaces can not be neglected. 

\begin{figure}[h]
 \begin{center}
  \includegraphics[width=220pt]{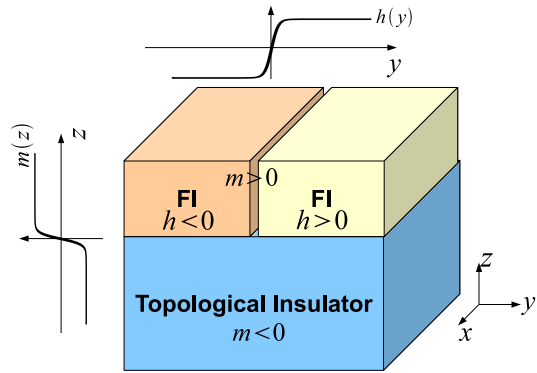}
 \end{center}
 \caption{(Color online) The heterostructure geometry for the topological insulator-ferromagnetic 
insulator(FI) tri-junction. }
 \label{fig3}
\end{figure}

For instance, the semiclassical approximation fails for a topological insulator-ferromagnet
heterostructure junction.
In the following, we demonstrate this by using
the three dimensional 
Dirac model for the surface of the topological 
insulator with ferromagnetic perturbation, 
\begin{equation}
\begin{split}
H = v \mu_1 \left( k_x \sigma_1 -i \partial_y \sigma_2 -i \partial_z \sigma_3 \right) + m(z) \mu _3 + h(y) \sigma_3.
\end{split}
\end{equation}
Here $\bm{\sigma} = (\sigma_1, \sigma_2, \sigma_3)$ and $\bm{\mu} = (\mu_1, \mu_2, \mu_3)$ 
are the Pauli matrices representing spin, and orbital degrees of freedom, respectively.
The mass parameter $m(z)$ describes the difference between trivial and topological phases.
$h(y)$ describe the domain wall of the ferromagnetic insulator.
We consider surface geometry in which $m>0$ for $z>0$ , $m<0$ for $z<0$, $h>0$ for $y>0$, and 
$h<0$ for $y<0$ as shown in FIG. \ref{fig3}.
The system has translational symmetry in the $x$-direction, 
and thus, $k_x$ is the only parameter of the Hamiltonian. 
This Hamiltonian possesses a chiral edge mode localized in the line defect at
$y,z \sim 0$. The associated wave function is 
\begin{equation}
\label{chiral_edge}
\begin{split}
&\Phi_{k_x}(y,z) \sim 
\begin{pmatrix}
1 \\
i \\
-i \\
-1
\end{pmatrix}
e^{-\int^{z} dz' m(z')/v } e^{-\int^{y} dz' h(y)/v } e^{i k_x x}, \\
&H \Phi_{k_x}(y,z) = -v k_x \Phi_{k_x}(y,z) . 
\end{split}
\end{equation}
up to the normalization factor.
Here we take the basis of spin and orbital degrees of freedom, 
$\left\{ \ket{\sigma_3=1,\mu_3=1} \right. $, $\left. \ket{\sigma_3=1,\mu_3=-1} \right.$, $ \left. \ket{\sigma_3=-1,\mu_3=1} \right. $, $\left. \ket{\sigma_3=-1,\mu_3=-1} \right\}$.
Within the semi-classical approximation, $\hat{\bm{p}} = -  i \bm{\nabla}$ and $\hat{\bm{x}}$ are treated as commutative variables. 
The semi-classical Hamiltonian describing the interface of topological insulator-ferromagnet is 
\begin{equation}
\begin{split}
&H(k_x,k_y,k_z;y,z) \\
& \ \ \ \ = v \mu_1 \left( k_x \sigma_1 + k_y \sigma_2 + k_z \sigma_3 \right) + m(z) \mu _3 + h(y) \sigma_3.
\end{split}
\end{equation}
The square of the energy eigenvalues are
\begin{equation}
\begin{split}
&E^2(k_x,k_y,k_z,y,z) \\
& \ \ \ \ = v^2 (k_x^2 + k_y^2) + \left( \sqrt{m^2(z) + v^2 k_z^2} \pm h(y) \right)^2.
\end{split}
\end{equation}
Near the interface between topological insulator and ferromagnet 
in which the condition $|m(z)|<|h(y)|$ is satisfied, 
there are the gap closing points at $(k_x,k_y,k_z) = (0,0,\pm \sqrt{h^2(y) - m^2(z)}/v)$. 
Therefor the second Chern number (\ref{second_Chern_number}) calculated from this incorrect spectrum cannot be quantized. 
This gap closing is a spurious one raised by the inaccurate semi-classical approximation.
To obtain the $\theta$-term for this junction system, we need quantum corrections which lead to
the correct energy spectrum.
In the next section, we construct the $\theta$-term in the full quantum manner.

As discussed in ref.[\onlinecite{TK,EMV,QHZ}], 
in the case that a time-reversal symmetry breaking term is a Dirac type $h_{af}(y,z) \mu_2$ expressed by $\gamma$-matrices :  
\begin{equation}
\begin{split}
H = -i v \mu_1 \bm{\sigma} \cdot \bm{\nabla} + m(y,z) \mu_3 + h_{af}(y,z) \mu_2, 
\end{split}
\label{H_antiferro}
\end{equation} 
the semi-classical approximation is successful, since the energy gap does not close. 
The five set of matrices $\{ \mu_1 \sigma_1, \mu_1 \sigma_2, \mu_1 \sigma_3, \mu_3, \mu_2 \}$ consist Dirac matrices 
$\gamma_0, \gamma_1, \gamma_2, \gamma_3$, and $\gamma_5$. 
Due to the anti-commutative property, 
the nonzero $m(y,z)$ and $h_{af}(y,z)$ play a role of a gap function, and 
set a characteristic length scale $\xi_c \sim v/\sqrt{m^2(y,z) + h^2_{af}(y,z)}$. 
To deal with the quantum corrections for the spatial inhomogeneity of $m(y,z)$ and $h_{af}(y,z)$, 
the gradient expansion can be applied as long as 
the spatial inhomogeneities are slower than $\xi_c$. 
Especially, the semi-classical approximation, which is the zeroth order of the gradient expansion, 
is also applicable. In fact, the semi-classical energy spectrum of the Hamiltonian (\ref{H_antiferro}),
\begin{equation}
\begin{split}
&E(k_x,k_y,k_z,y,z) \\
&\ \ \ \ = \pm \sqrt{v^2 (k_x^2 + k_y^2 + k_z^2) + m^2(y,z) + h_{af}^2(y,z)}, 
\end{split}
\end{equation}
has an energy gap unless $m^2(y,z) + h_{af}^2(y,z) = 0$, i.e. away from the defect line, 
and hence, the second Chern number (\ref{second_Chern_number}) is well defined. 

On the other hand, in the case of the topological insulator-ferromagnet junction, the Zeeman term $h \sigma_3$, which 
is not expressed in terms of $\gamma$ matrices, 
does not protect the energy gap 
since the $h \sigma_3$ commute with $k_z \mu_1 \sigma_3$. 
Thus the gradient expansion and semi-classical approximation cannot be applied to the interface of topological insulator-
ferromagnet.

\section{The Construction of the Topological Invariant for line defects with broken time-reversal symmetry}
\label{sec3}

In this section, we present the full quantum treatment of the topological number characterizing line defects without time reversal symmetry in three dimensional Insulators. 
This can be achieved by using the Green's function.
In spatially inhomogeneous systems, the single particle Green's functions 
$\tilde G_{\alpha \beta}(i \omega , \bm{x}_1, \bm{x}_2)$ is defined by the Dyson's equation, 
\begin{equation}
\begin{split}
\int d\bm{x}_2 \tilde K (i \omega, \bm{x}_1, \bm{x}_2) \tilde G(i \omega , \bm{x}_2, \bm{x}_
3) = \delta (\bm{x}_1 - \bm{x}_3) .
\end{split}
\end{equation}
$\tilde K (i \omega, \bm{x}_1, \bm{x}_2)$ is the kernel of the Green's function. 
For noninteracting systems $\tilde K (i \omega, \bm{x}_1, \bm{x}_2) = \left[ i \omega - H(-i \bm{\nabla}_1, \bm{x}_1) \right] \delta (\bm{x}_1 - \bm{x}_2)$.
In this cases, Dyson's equation is 
\begin{equation}
\begin{split}
\left[ i \omega - H(-i \bm{\nabla}_1, \bm{x}_1) \right] \tilde G(i \omega , \bm{x}_1, \bm{x}_
2) = \delta (\bm{x}_1 - \bm{x}_2) .
\end{split}
\end{equation}
Then we consider the Wigner transformation of $\tilde G(i \omega , \bm{x}_1, \bm{x}_2)$, 
\begin{equation}
\begin{split}
G(i \omega , \bm{p}, \bm{R}) 
:= \int d \bm{r} \ \tilde G \left(i \omega , \bm{R}+ \frac{\bm{r}}{2}, \bm{R}- \frac{\bm{r}}{2} \right) e^{-i \bm{p} \cdot \bm{r}} .
\end{split}
\end{equation}
Here, $\bm{p}$ is the momentum of relative coordinate $\bm{r} = \bm{x}_1-\bm{x}_2$,
 and $\bm{R} = (\bm{x}_1+\bm{x}_2)/2$ is the center of mass coordinate. 
For translational symmetric systems, $G$ is the matrix inverse of $K$, $G=K^{-1}$, 
but in the case of inhomogeneous systems, $G \neq K^{-1}$.
We introduce a
closed path $C = \bm{R}(s)$ surrounding the line defect in real space.
If the Green's function $G(i \omega, \bm{p}, \bm{R})$ is not singular for all $\omega$, $\bm{p}$ 
on the closed path $C$, then topological invariant $N$ can be defined by 
\begin{equation}
\begin{split}
N =& -\frac{i}{480 \pi^3} \oint_{C} ds \int_{-\infty}^{\infty} d \omega  \int d^3\bm{p} \ \epsilon^{\mu \nu \rho \sigma \eta} \\
& \ \ \ \ \ \mathrm{tr} \left[ G^{-1} \partial_{\mu} G G^{-1} \partial_{\nu} G G^{-1} \partial_{\rho} G G^{-1} \partial_{\sigma} G G^{-1} \partial_{\eta} G  \right], 
\end{split}
\label{topologicalinvariant}
\end{equation}
where $(\mu, \nu, \rho, \sigma, \eta)$ run over $(\omega, p_x, p_y, p_z, s)$, $\epsilon$ is fully 
anti-symmetric tensor with $\epsilon^{\omega p_x p_y p_z s} = 1$, and 
$\partial_s = \frac{d \bm{R}(s)}{d s} \cdot \frac{\partial}{\partial \bm{R}}$ is the directional 
derivative along the path $C$. 
This topological number characterize the homotopy of the map 
$G : (\omega, p_x, p_y, p_z, s) \mapsto G(i\omega, p_x, p_y, p_z, s) \in \mathrm{GL}(n,\mathbb{C})$, 
here $n$ is the number of the band. 
If the base space $(\omega, p_x, p_y, p_z, s)$ can be considered as $S^5$ 
this map is classified by the homotopy group 
$\pi_5\left( \mathrm{GL}(n,\mathbb{C}) \right) = \mathbb{Z}$, 
which imply the existence of the nontrivial topological line defects. 
The same topological invariant as (\ref{topologicalinvariant}) was previously obtained by Silaev and Volovik \cite{SV} 
by using gradient expansions for the semi-classical 
Green's function. 
We stress that, in our approach, we have constructed the topological invariant (\ref{topologicalinvariant}) from
the argument on the homotopy of the exact Wigner-transformed Green's function, without using gradient expansion, and thus
Eq.(\ref{topologicalinvariant}) is a full-quantum expression free from semi-classical approximations. 

The topological invariant  
$N$ is not changed by the deformation of the closed path $C$ 
as long as the singularity does not occur on the path $C$. 
The nontrivial topological invariant $N \neq 0$ means the existence of 
the singularity of $G$ inside $C$, say, on the line defect. 
The singularity of Green's function $G$ arises from the existence of ill-defined points in the map $G$, 
which correspond to
$0$ or $\infty$ eigenvalues of the matrix $G(i\omega,\bm{p},\bm{R})$ 
at some points $(\omega,\bm{p})$ on the line defect coordinate $\bm{R}$. 
The $\infty$ eigenvalues imply the existence of the gapless modes localized 
at the line defects because only $\omega=0$ conforms to the $\infty$ eigenvalue for the definition
of $G$. Also, as noted by [\onlinecite{Gurarie,EG}], in the case with electron correlation effects, 
the $0$ eigenvalue is allowed, and associated with the Mott insulating phase.

 
The topological invariant $N$ can be represented as the winding number of a 
certain potential function $\theta(\bm{R})$. 
This can be seen by rewriting eq.(\ref{topologicalinvariant}) as
\begin{equation}
\begin{split}
&N=\frac{1}{2 \pi} \oint_{C} d\bm{R}(s) \cdot \bm{A}(\bm{R}),  \\
&\bm{A}(\bm{R}) = -\frac{i}{48 \pi^2} \int_{-\infty}^{\infty} d \omega  \int d^3\bm{p} \ \epsilon^{\mu \nu \rho \sigma} \\ 
& \ \ \ \ \mathrm{tr} \left[ G^{-1} \partial_{\bm{R}} G G^{-1} \partial_{\mu} G G^{-1} \partial_{\nu} G G^{-1} \partial_{\rho} G G^{-1} \partial_{\sigma} G  \right] .
\end{split}
\end{equation}
According to the path-independent property of the line integral, $\bm{A}(\bm{R})$ is rotation free, 
$\bm{\nabla} \times \bm{A} = \bm{0}$. 
Hence $\bm{A}(\bm{R})$ can be represented as the gradient 
of a certain potential function $\theta(\bm{R})$, $\bm{A}(\bm{R}) = \bm{\nabla} \theta(\bm{R})$. 
Due to the Hermiticity of the Hamiltonian, $\bm{A}(\bm{R})$ is real, so $\theta(\bm{R})$ is also the 
real function. 
Thus we get 
\begin{equation}
\begin{split}
N =\frac{1}{2 \pi} \oint_{C} d\theta(\bm{R}). 
\end{split}
\end{equation}

In the cases of the slow limit of spatial variation or within the semi-classical approximation, 
the Green's function is reduced to a simple form, 
$G(i\omega,\bm{p},\bm{R}) = \left[ i \omega - H(\bm{p}, \bm{R}) \right]^{-1}$. 
Then $N$ is equal to the 2nd Chern number of the semi-classical Hamiltonian 
$H(\bm{p},s) := H(\bm{p},\bm{R}(s))$: 
\begin{equation}
\begin{split}
N = Ch_2.
\end{split}
\end{equation}
in consistent with the Teo-Kane results.\cite{TK}

In the semi-classical approximation, the above expression of the $\theta$ in terms of the Green's function is the same as the formula of the $\theta$ term introduced by Qi, Hughes, and Zhang for the Axion electrodynamics of the topological insulators.\cite{QHZ}
It is noted that eq. (\ref{topologicalinvariant}) represents the topological invariant even when
the semiclassical approximation, and the gradient expansion for the derivation of the Axion electrodynamics action are not justified, as in the case of the topological insulator-ferromagnet junction mentioned in Sec. II.
However, since we do not know how to derive the Axion electrodynamics action without using the gradient expansion,\cite{QHZ,EMV}  we could not make a direct connection 
between the $\theta$ introduced above and the $\theta$-term
of the Axion electrodynamics in the case that the semiclassical approximation is not applicable. 

Here, we make a general remark on the calculations of topological invariant associated with defects. 
To avoid the singularity at the defects, the closed line or surface surrounding the defects 
must be enough far away from the defects.
For example, in the case of line defects in heterostructure systems with broken time reversal symmetry, 
we calculate the changes of the $\theta$-term on the interfaces of insulators which are located far 
away from line defects by using eq.(11). 
This procedure is the same as the calculation of the bulk topological invariants
for which effects of boundaries are neglected.

\section{An Exactly Solvable Model for a Topological insulator-ferromagnet junction}
\label{sec4}

In this section and the next section, we demonstrate that
the formula (\ref{topologicalinvariant}) gives the correct topological invariant even
for the case where the semi-classical approach fails.
For this purpose, we use an exactly solvable model of
a topological insulator-ferromagnet heterostructure junction, which allows us the full quantum treatment
of the spatially inhomogeneous system. 
We will calculate the variation of $\theta$ on the surface of this system. 
The model Hamiltonian has a kink structure in the mass parameter $m$, which distinguish between 
topological and trivial phases,
\begin{equation}
\label{Hamiltonian_TI-FM}
\begin{split}
H = v \mu_1 \left( k_x \sigma_1 + k_y \sigma_2 -i \partial_z \sigma_3 \right) + m(z) \mu _3 + h \sigma_3.
\end{split}
\end{equation}
We consider surface geometry in which $m>0$ for $z>0$, and $m<0$ for $z<0$. 
First we investigate the energy spectrum of this Hamiltonian. 
According to the appendix \ref{app2}, the energy spectrum of this Hamiltonian 
(\ref{Hamiltonian_TI-FM}) can be labeled by the using of 
the eigenvalues $\epsilon$ of the Jackiw-Rebbi Hamiltonian, 
$H^{\uparrow }_2(z) := -i \mu_1 \partial_z + m(z) \mu _3$. 
The results are 
\begin{equation}
\begin{split}
E_{\epsilon} (k_x,k_y) = &\pm \sqrt{ (\epsilon + h)^2 + v^2 (k_x^2 + k_y^2)}, \\
&\pm \sqrt{ (\epsilon - h)^2 + v^2 (k_x^2 + k_y^2) },
\end{split}
\end{equation}
for $\epsilon > 0$, and 
\begin{equation}
\begin{split}
E_{0} (k_x,k_y) = \pm \sqrt{ v^2 (k_x^2 + k_y^2)+h^2},
\end{split}
\end{equation}
for $\epsilon = 0$. 
As shown in Appendix \ref{app3}, in the case of $m(z) = m \tanh (z/\xi)$,
we can obtain the exact Green function as well as the exact energy spectrum.
\cite{SK,TLM}
The possible values of $\epsilon$ are classified into the bound states $\epsilon^B$ localized at $z\sim 0$, and the continuum states 
$\epsilon^C$ 
\begin{equation}
\begin{split}
&\epsilon^{B}_0 = 0 ,  \\
&\epsilon^{B}_{n,\pm} = \pm m \sqrt{\frac{n}{\nu} \left( 2-\frac{n}{\nu} \right)} \ , (n = 1,2,\cdots <\nu )\ , \\
&\epsilon^{C}_{p,\pm} = \pm \sqrt{m^2 + v^2 p^2}\ , 
\end{split}
\end{equation}
where $\nu = \frac{m \xi}{v}$ describing the spatial scale of the kink. 
When $\nu > 1$, bound states with nonzero energies are allowed
as well as
a zero energy state, which corresponds to $\epsilon^{B} = 0$. 
The energy dispersion of the bound states 
\begin{equation}
\begin{split}
E_{n,\pm} (k_x,k_y) &= \pm \sqrt{ (\epsilon_{n \pm} - h)^2 + v^2 (k_x^2 + k_y^2)} , \\ 
& \hspace{90pt} (n = 1,2,\cdots <\nu )\ 
\end{split}
\end{equation}
implies that the energy gap is closed only when $| \epsilon_{n \pm} | = | h |$. 
This property is consistent with the gap closing found in the semi-classical approximation
because of the following reason. 
The semi-classical approximation corresponds to the infinitely slow varying kink potential : 
$\xi = \infty $. Then the bound states distribute in the range of $-m < \epsilon_n < m$ densely, 
and hence the energy gap is closed because there are the bound states with
$| \epsilon_{n \pm} | \sim | h |$. 
This behavior is drastically changed in the cases of the topological insulator-antiferromagnet 
heterostructure considered in ref.[\onlinecite{TK}]. According to the Appendix \ref{app2}, in the case of the anti-ferromagnet, 
the energy dispersion for the bound states are 
\begin{equation}
\begin{split}
E^{\mathrm{AF}}_{n,\pm} (k_x,k_y) &= \pm \sqrt{ \epsilon^2_{n \pm} + h^2_{af} + v^2 (k_x^2 + k_y^2)} , \\ 
& \hspace{90pt} (n = 1,2,\cdots <\nu ), \ 
\end{split}
\end{equation}
The system is the fully gapped with respect to the arbitrary large $\xi$, 
so the semi-classical approximation is valid. 

In real materials, $\xi$ is atomic scale, and continuum approximation is not valid for describing 
the interface structures. However, we are mainly concerned with how the $\theta$ term changes at the interfaces. 
This change is a topological property, and does not depend on the details of the microscopic structure of the interfaces as 
long as the energy gap is not closed. 
Thus, for simplicity, in the following, we set $\nu = 1 \Leftrightarrow \xi = v/m$,
for which the expression of the exact Green's function is considerably simplified. 

Here, we summarize the exact results for the Green function of the model (\ref{Hamiltonian_TI-FM}).
The detail of the derivation is presented in Appendixes \ref{app2.5}, \ref{app3}, and \ref{app4}.
It is convenient to
introduce the polar coordinates with respect to $(\omega, v k_x, v k_y)$, $ \omega = \rho \sin \theta , v k_x = \rho \cos \theta \cos \phi ,  v k_y = \rho \cos \theta \sin \phi$. 
Then, the Green's function is decomposed as 
\begin{equation}
\label{pd_G}
\begin{split}
&G(\rho,\theta,\phi;z_1,z_2) \\
& \ \ \ \ = U^{\dag}(\phi) V^{\dag}(\theta) W^{\dag} \tilde G(\rho,z_1,z_2) W \sigma_2 V(\theta) \sigma_2 U(\phi) ,
\end{split}
\end{equation}
where 
\begin{equation}
\begin{split}
U(\phi) = e^{\frac{i}{2}  \sigma_3\phi} ,
V(\theta) = e^{\frac{i}{2} \mu_1 \sigma_1 \theta} ,
W = \frac{1}{\sqrt{2}} (1 + i \mu_1), 
\end{split}
\end{equation}
and 
$\tilde G(\rho,z_1,z_2)$ is the Green's function for the one-dimensional Hamiltonian $\tilde H(z)$ : 
\begin{equation}
\begin{split}
\tilde H(z) = -i v \partial_z \mu_1  \sigma_3 + m(z) \mu_2 + h \sigma_3.
\end{split}
\end{equation}
$\tilde H(z)$ is diagonal in the spin space, $\tilde H(z) = \tilde H^{\uparrow }(z) \oplus \tilde H^{\downarrow }(z)$.
Then the problem to calculate the Green's function reduces to solving one-dimensional 
Hamiltonians in the orbital space, 
\begin{equation}
\begin{split}
&\tilde H^{\uparrow}(z) = -i v \partial_z \mu_1 + m \tanh \left( \frac{z}{\xi} \right) \mu_2 +h, \\
&\tilde H^{\downarrow}(z) = i v \partial_z \mu_1 + m \tanh \left( \frac{z}{\xi} \right) \mu_2 -h. 
\end{split}
\end{equation}
If $h=0$ these systems are equivalent to the Jackiw-Rebbi model\cite{JR} in the one-dimensional chiral symmetric system.
The Jackiw-Rebbi model possesses a zero energy soliton solution, and the finite $h$ induces 
the shift of the energy. 
The soliton modes in the two sectors $\tilde H^{\uparrow }(z)$ and $\tilde H^{\downarrow }(z)$ 
correspond to the helical Dirac fermions on the surface of the topological insulators. 

When $\nu=1$ the eigenfunctions for $\sigma_3 = \ \uparrow $ sector take simple analytical forms:
\begin{equation}
\begin{split}
&\Phi^{\uparrow }_{0} (z) = \sqrt{\frac{m}{2 v}} 
\begin{pmatrix}
0 \\
1
\end{pmatrix} 
\mathrm{sech} \left( \frac{m}{v} z \right) , \\
&\Phi^{\uparrow }_{p,\pm} (z) = \frac{1}{\sqrt{2}}
\begin{pmatrix}
1 \\
\frac{v p + i m \tanh (\frac{m}{v} z)}{\pm \sqrt{v^2 p^2 + m^2}}
\end{pmatrix} 
e^{i p z} ,
\end{split}
\end{equation}
with eigenvalues $\epsilon_0 = h$, and $\epsilon_{p,\pm} = \pm \sqrt{m^2 + v^2 p^2} + h$ respectively. 
Due to the completeness property of the set of eigenfunctions 
$\left\{ \Phi^{\uparrow }_{0}, \Phi^{\uparrow }_{p,\pm} \right\}$, 
the Green's function $\tilde G^{\uparrow }(\rho,z_1,z_2) $ can be expressed by 
the spectral representation as follows,
\begin{equation}
\label{G(z)}
\begin{split}
& \tilde G^{\uparrow }(i \rho , z_1 , z_2) \\
& \ \ \ \ = \sum_{\pm} \int \frac{d p}{2 \pi} \ \frac{\Phi^{\uparrow }_{p,\pm} (z_1) \left[ \Phi^{\uparrow }_{p,\pm}(z_2) \right]^{\dag} }{i \rho - \epsilon_{p,\pm}} + \frac{\Phi^{\uparrow }_0 (z_1) \left[ \Phi_0^{\uparrow }(z_2)\right]^{\dag} }{i \rho - \epsilon_0} .
\end{split}
\end{equation}
The Green's function of $\sigma_3 = \ \downarrow $ sector is obtained in a similar way.
The Wigner transformation of $\tilde G^{\uparrow }$, and $\tilde G^{\downarrow }$ is given by,
\begin{widetext}
\begin{equation}
\label{exact_GF}
\begin{split}
& G^{\uparrow }(i\rho,p,R) = \frac{1}{(i \rho-h)^2-v^2 p^2-m^2}
\begin{pmatrix}
i \rho-h & v p - i \tilde m (\rho,-v p,R/v;h) \\
v p + i \tilde m (\rho,v p,R/v;h) & (i \rho - h) \beta (\rho,v p,R/v;h)
\end{pmatrix}, \\
& G^{\downarrow }(i\rho, p,R) = \frac{1}{(i \rho+h)^2-v^2 p^2-m^2}
\begin{pmatrix}
(i \rho + h) \beta (\rho,v p,R/v;-h) & -p - i \tilde m (\rho,v p,R/v;-h) \\
-v p + i \tilde m (\rho,-v p,R/v;-h) & i \rho+h
\end{pmatrix}.
\end{split}
\end{equation}
\end{widetext}
The derivation of (\ref{exact_GF}) and the definition of the effective mass $\tilde m$, 
and the factor $\beta$ are presented in Appendix \ref{app4}. 

After integrating over
$\theta$, and $\phi$, we obtain (see Appendix \ref{app5} for details),
\begin{equation}
\begin{split}
A_z(R_z) 
&= \frac{i}{48 \pi} \int_0^{\infty} d\rho \int dp_z \ \epsilon^{\mu \nu \rho \sigma } \\ 
& \ \ \ \ \ \ \ \ \ \ \ \ \ \ \ \ \mathrm{tr} \left[ \mu_1 \sigma_2 \left\{ \chi_{\mu}\chi_{\nu}\chi_{\rho}\chi_{\sigma} + \xi_{\mu}\xi_{\nu}\xi_{\rho}\xi_{\sigma} \right\} \right] \ ,  
\end{split}
\label{A(R)}
\end{equation}
where 
\begin{equation}
\label{chi_xi}
\begin{split}
\chi_{\mu} (\rho,p_z,R_z) &=  \tilde G^{-1} \partial_{\mu} \tilde G \ \ \ \ (\mu \in (\rho,p_z,R_z)) \ , \\
\chi_{\theta}(\rho,p_z,R_z) &= \mu_1 \sigma_1 + \tilde G^{-1} \mu_1 \sigma_1 \tilde G \ , \\
\xi_{\mu}(\rho,p_z,R_z)&= \partial_{\mu} \tilde G G^{-1} \ \ \ \ (\mu \in (\rho,p_z,R_z)) \ , \\
\xi_{\theta}(\rho,p_z,R_z) &= \mu_1 \sigma_1 + \tilde G \mu_1 \sigma_1 \tilde G ^{-1}\ ,
\end{split}
\end{equation}
$\tilde G = \tilde G^{\uparrow } \oplus \tilde G^{\downarrow }$, and $\epsilon$ is fully anti-symmetric 
tensor with $\epsilon^{\rho \theta p_z R_z} = 1$.
The derivation of Eq . (\ref{A(R)}) is in Appendix \ref{app5}.
Eqs. (\ref{A(R)}) and (\ref{chi_xi}) are used for the numerical calculation of $\theta$ and the topological invariant (\ref{topologicalinvariant})
in the next section.

\section{Numerical Results of $\theta$ and the topological invariant for the topological insulator-ferromagnet junction}
\label{sec5}

\begin{figure}[h]
 \begin{center}
  \includegraphics[width=220pt]{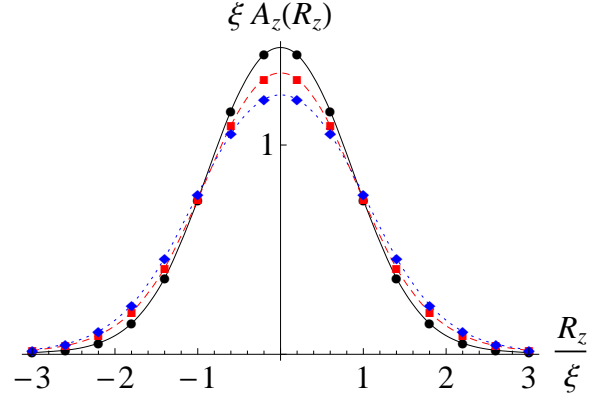}
 \end{center}
 \caption{(Color online) The numerical calculations of $A_z(R_z)$ given by the Hamiltonian 
(\ref{Hamiltonian_TI-FM_2}). 
Points in figure are the results of numerical calculation for sample points $R_z$. 
The smooth lines are given by interpolation of the points.
Black (solid), red (dashed), and blue (dotted) lines correspond to $h=0.1 m$, $0.5 m$, and $0.9 m$, 
respectively.
}
 \label{fig1}
\end{figure}

\begin{figure}[h]
 \begin{center}
  \includegraphics[width=220pt]{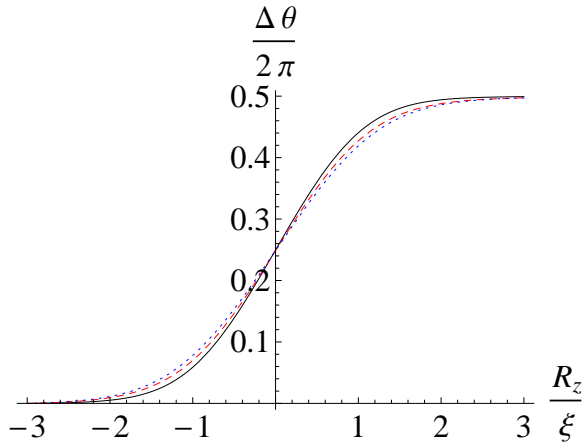}
 \end{center}
 \caption{(Color online) The variations of $\theta$ on the interface of topological insulator-
 ferromagnet heterostructure.
 Black (solid), red (dashed), and blue (dotted) lines correspond to $h=0.1 m$, $0.5 m$, and $0.9 m$, 
respectively.}
 \label{fig2}
\end{figure}

In this section, we present the results of numerical calculations of the variation of $\theta$, $\Delta \theta$, in the vicinity of interfaces of 
the topological insulator-ferromagnet tri-junction system, for which the semi-classical approximation
is not applicable. 
The heterostructure system considered here consists of the topological insulator and
the two ferromagnets with the opposite direction of the magnetization.
There are three interfaces; i.e.
one between the topological insulator and the ferromagnet with the internal magnetic field $h>0$
(interface I),
another between the topological insulator and the ferromagnet with $h<0$ (interface II), and the other
between the two ferromagnets with $h>0$ and $h<0$ (interface III).
The intersection of these three interfaces forms a line defect.
The Hamiltonian in the vicinity of the interface I or II is,
\begin{equation}
\label{Hamiltonian_TI-FM_2}
\begin{split}
&H(k_x,k_y,z;h) \\
& \ \ = v \mu_1 \left( k_x \sigma_1 + k_y \sigma_2 - i \partial_z \sigma_3 \right) + m \tanh (z/\xi) \mu _3 + h \sigma_3. 
\end{split}
\end{equation}
Here we set $\frac{m \xi}{v} = 1$. 
First, we consider the interface I ($h>0$).
In Fig.\ref{fig1}, we show the numerical results of $A_z(R_z)$ for this model
calculated by using (\ref{exact_GF}), (\ref{A(R)}), and (\ref{chi_xi})
for three values of $h>0$: $h=0.1 m$, $0.5 m$, and $0.9 m$. 
Also, we show
in Fig.\ref{fig2} the variations of $\theta$ obtained from the integration of 
$A_z(R_z)$, $\Delta \theta (R_z) = \int_{-\infty}^{R_z} A(R_z') d R_z'$.
It is found that for each value of $h$, the total change of $\theta$ is $\pi$. 
This results are consistent with the $\theta_{\mathrm{bulk}}$ defined by the Chern-Simons 3 form. 
Here we introduce the notation "$\theta_{\mathrm{bulk}}$" to distinguish from the $\theta$ 
defined as the potential of 
$\bm{A}$. 
The $\theta_{\mathrm{bulk}}$ describes the bulk structures of insulators. 
Since inversion symmetry is not broken in the bulk ferromagnet and the bulk topological insulator,
the value of $\theta_{\mathrm{bulk}}$ is fixed to $0$ for trivial insulators, 
or $\pi$ for topological insulators. 
Our method provides a reliable method for calculating  the spatial change of $\theta$ which 
connects the trivial value $0$ in the ferromagnet region and the nontrivial value $\pi$ in the topological insulator region.


Next we consider the interface II ($h<0$). 
Just at $h=0$, the induced mass gap in the helical Dirac fermion on the surface of topological 
insulator disappears. 
Then the total change of $\theta$ is allowed to be a different value from the case of $h>0$ 
since the Green's function have singularities when the energy spectrum is closed. 
In fact, the relation, 
\begin{equation}
\begin{split}
\mu_3 \sigma_1 H(k_x,k_y,z;h) \mu_3 \sigma_1 = H(-k_x,k_y,z;-h), 
\end{split}
\end{equation}
follows 
\begin{equation}
\begin{split}
&\mu_3 \sigma_1 G(i \omega, k_x, k_y, p_z, R_z; h) \mu_3 \sigma_1 \\
& \ \ \ \ \ \ \ \ \ \ = G(i \omega, -k_x, k_y, p_z, R_z; -h), 
\end{split}
\end{equation}
and 
\begin{equation}
\label{A_z(h)}
\begin{split}
A_z(R_z;h) = - A_z(R_z;-h). 
\end{split}
\end{equation}
Hence the total change of $\theta$ for $h<0$ is $-\pi$. 
The sign of the change of $\theta$ is determined by the sign of $h$, i.e. 
the sign of the induced mass gap.\cite{QHZ} 

Finally, we consider the variation of $\theta$ on the domain wall of ferromagnet (the interface III),
the Hamiltonian of which is given by 
\begin{equation}
\label{Hamiltonian_FM-FM}
\begin{split}
&H(k_x,y,k_z) \\
& \ \ = v \mu_1 \left( k_x \sigma_1 - i \partial_y \sigma_2 + k_z \sigma_3 \right) + m \mu _3 + h(y) \sigma_3. 
\end{split}
\end{equation}
Here $h(y)$ describes the domain wall structure of ferromagnet. 
it is found that $\theta$ does not vary on the domain wall of ferromagnet since 
the relation, 
\begin{equation}
\begin{split}
\mu_3 \sigma_3 H(k_x,y,k_z) \mu_3 \sigma_3 = H(k_x,y,-k_z),  
\end{split}
\end{equation}
follows 
\begin{equation}
\begin{split}
&\mu_3 \sigma_3 G(i \omega, k_x, p_y, k_z, R_y) \mu_3 \sigma_3 \\
& \ \ \ \ \ \ \ \ \ \ = G(i \omega, k_x, p_y, -k_z, R_y), 
\end{split}
\end{equation}
and 
\begin{equation}
\begin{split}
A_y(R_y) = -A_y(R_y) = 0.
\end{split}
\end{equation}

The total change of $\theta$ about the closed path $C$ surrounding the line defect in 
topological insulator-ferromagnet tri-junction is $\pi + 0 + \pi = 2 \pi$, 
then we get the topological invariant $N = 1$. 
The nontrivial topological invariant means the existence of 
the singularity of $G$ on the line defect. 
Then our result is consistent with the existence of chiral edge mode described by 
eq. (\ref{chiral_edge}).

\section{Discussions and Conclusion}
\label{sec6}
In this paper we proposed a full quantum formulation for the topological invariant 
characterizing line defects in three-dimensional insulators with no symmetry by using 
the Green's function method. 
The nontrivial topological invariant leads to the existence of gapless modes localized in 
the line defects. 
Our approach is applicable to various heterostructure systems involving topological insulators,
including the case that a semiclassical approximation for spatial inhomogeneity
raised by defects fails to describe topological features of the systems.
As an example of the failure of a semiclassical approximation, we considered the topological insulator-ferromagnet junction system, which is important for the application to the quantum Hall effect.
To deal with the issue beyond semi-classical approximations, we employed
an exactly solvable model of the heterostructure junction, for which the exact Green's function can be
obtained.
On the basis of the Green's function representation of the topological invariant, we demonstrated the nontrivial topological invariant for the  topological insulator-ferromagnet 
heterostructure junction, which
implies the existence of a chiral edge mode leading to the quantum Hall effect. 
Our method provides a reliable method for calculating  the spatial change of $\theta$ which 
connects the trivial value $0$ in the ferromagnet region and the nontrivial value $\pi$ in the topological insulator region.

Within the semi-classical approximation, our formula of the
topological invariant for line defects and the associated $\theta$ term expressed
 in terms of the Green's function are, respectively, the same as 
 the Teo-Kane's formula and the Qi-Hughes-Zhang's formula for the $\theta$ term of the Axion electrodynamics.
Since we do not know how to derive the Axion electrodynamics action without using the gradient expansion, we could not make a direct connection 
between the $\theta$ introduced in this paper and the $\theta$-term
of the Axion electrodynamics in the case that the semiclassical approximation is not applicable. 
We believe that even in such a case, the $\theta$ expressed in terms of the Green's function 
is related to electromagnetic response functions characterizing the Axion electrodynamics of the topological insulators.
The clarification of this point should be addressed in the near future.

There are several future directions of the current study.
For example, our approach based on the Green's function may be useful for the study on effects of electron-electron interactions and impurity scattering in the topological insulator with spatial inhomogeneity.
Another direction is the application 
to the case of carrier-doped topological insulators, in which bulk metallic states coexist
with surface gapless states, as realized in various three-dimensional topological insulators.\cite{HQW,CHEN,XWQ}
Recently, the Axion response in such gapless systems has been discussed.\cite{bergman,BQI}
For such a case, effects of impurity scattering on the Axion response involving the bulk gapless states
may become important, and can be evaluated by using our Green's function method.

\begin{acknowledgments}

This work is supported by the Grant-in-Aids for
Scientific Research from MEXT of Japan (Grants 
No. 19052003, No. 21102510, and No. 23540406).

\end{acknowledgments}

\onecolumngrid

\appendix

\section{The Energy Spectrum of the Hamiltonian (\ref{Hamiltonian_TI-FM})}
\label{app2}

Hamiltonian (\ref{Hamiltonian_TI-FM}) is decomposed into two parts:
\begin{equation}
\begin{split}
&H(k_x,k_y;z) = H_1(k_x,k_y) + H_2(z) \\
&H_1(k_x,k_y) = v \mu_1 (k_x\sigma_1 +k_y \sigma_2) + h \sigma_3 \\
&H_2(z) = - i v \mu_1 \partial_z \sigma_3 + m(z) \mu_3 .
\end{split}
\end{equation}
$H_2(z)$ is diagonal in the spin space, $H_2 (z) = H^{\uparrow }_2(z) \oplus H^{\downarrow }_2(z)$. 
Due to the symmetric properties of $H_2(z)$ : $\mu_2 H_2 (z) \mu_2 = -H_2(z)$, 
and $H^{\downarrow}_2(z) = \mu_3 H^{\uparrow }_2(z) \mu_3$, 
the eigenfunctions of $H_2(z)$ with nonzero eigenvalue 
can be represented as a set of four eigenfunctions with energy 
$\epsilon,-\epsilon, \epsilon, -\epsilon$ : 
\begin{equation}
\begin{split}
\Phi_{\epsilon}(z) = 
\left\{
\begin{pmatrix}
\chi_{\epsilon}(z) \\
0 
\end{pmatrix}
,
\begin{pmatrix}
-i \mu_2 \chi_{\epsilon}(z) \\
0 
\end{pmatrix}
,
\begin{pmatrix}
0 \\
\mu_3 \chi_{\epsilon}(z)
\end{pmatrix}
,
\begin{pmatrix}
0 \\
\mu_1 \chi_{\epsilon}(z)
\end{pmatrix}
\right\}\ , \ \ \ (\epsilon > 0)\ , 
\end{split}
\end{equation} 
here $\chi_{\epsilon}(z)$ is the eigenfunction of $H^{\uparrow}_2(z)$ with the eigenvalue $\epsilon$ : 
$H^{\uparrow}_2(z) \chi_{\epsilon}(z) = \epsilon \chi_{\epsilon}(z)$.
Then $H(k_x,k_y;z)$ is closed in the subspace spanned by $\Phi_{\epsilon}(z)$, 
\begin{equation}
\begin{split}
H(k_x,k_y,z) \Phi_{\epsilon}(z) = \Phi_{\epsilon}(z) \left[ v \mu_1 (k_x\sigma_1 +k_y \sigma_2) + h \sigma_3 + \epsilon \mu_3 \right] .
\end{split}
\end{equation}
By diagonalizing of $v \mu_1 (k_x\sigma_1 +k_y \sigma_2) + h \sigma_3 + \epsilon \mu_3$, 
we get the energy spectrum labeled by $\epsilon$, 
\begin{equation}
\begin{split}
E_{\epsilon} (k_x,k_y) = \pm \sqrt{ (\epsilon + h)^2 + v^2 (k_x^2 + k_y^2)}\ \ , \ \ \pm \sqrt{ (\epsilon - h)^2 + v^2 (k_x^2 + k_y^2) }.
\end{split}
\end{equation}
Furthermore, $H_2(z)$ have the pair of zero energy bound states, 
\begin{equation}
\begin{split}
\Phi_{0}(z) = 
\left\{
\begin{pmatrix}
1 \\
i \\
0 \\
0 
\end{pmatrix}
,
\begin{pmatrix}
0 \\
0 \\
1 \\
-i
\end{pmatrix}
\right\}
e^{-\int^z dz' m(z')/v} .
\end{split}
\end{equation} 
$H(k_x,k_y,z)$ is closed in the subspace spanned by $\Phi_{0}(z)$, 
\begin{equation}
\begin{split}
H(k_x,k_y,z) \Phi_{0}(z) = \Phi_{0}(z) \left[ v (k_x\tau_2 -k_y \tau_1) + h \tau_3 \right] .
\end{split}
\end{equation}
Here the Pauli matrices $\bm{\tau}$ are defined in the subspace spanned by $\Phi_{0}(z)$. 
By diagonalizing of $v (k_x\tau_1 +k_y \tau_2) + h \tau_3$, 
we get the energy spectrum for zero energy bound states of $H_2(z)$, 
\begin{equation}
\begin{split}
E_0(k_x,k_y) = \pm \sqrt{v^2 (k_x^2 + k_y^2) + h^2 } . 
\end{split}
\end{equation}
The eigenfunctions $\Phi_{0}(z)$ corresponds to the degrees of freedom for helical Dirac fermion 
on the surface of topological insulator.

Here we remark the cases of topological insulator-antiferromagnet heterostructure described by the Hamiltonian considered in ref.[\onlinecite{TK}], 
\begin{equation}
\begin{split}
H^{\mathrm{AF}}(k_x,k_y;z) = v \mu_1 (k_x\sigma_1 +k_y \sigma_2) + h_{af} \mu_2 - i v \mu_1 \partial_z \sigma_3 + m(z) \mu_3 .
\end{split}
\end{equation}
Then the Hamiltonian $H^{\mathrm{AF}}(k_x,k_y;z)$ is closed in the same subspace spanned by $\Phi_{\epsilon}(z)$, 
\begin{equation}
\begin{split}
H^{\mathrm{AF}}(k_x,k_y,z) \Phi_{\epsilon}(z) = \Phi_{\epsilon}(z) \left[ v \mu_1 (k_x\sigma_1 +k_y \sigma_2) + h_{af} \mu_2 + \epsilon \mu_3 \right] .
\end{split}
\end{equation}
The energy spectrum energy spectrum labeled by $\epsilon$ is 
\begin{equation}
\begin{split}
E^{\mathrm{AF}}_{\epsilon} (k_x,k_y) = \pm \sqrt{ \epsilon^2 + h_{af}^2 + v^2 (k_x^2 + k_y^2)} .
\end{split}
\end{equation}
Here the upper and lower bands are doubly degenerated, respectively. 
The energy spectrum for zero energy bound states is the same form for topological insulator-ferromagnet 
heterostructure.

\section{The Polar Decomposition of Green's Function}
\label{app2.5}
The kernel of the Green's function associated with Hamiltonian (\ref{Hamiltonian_TI-FM}), 
\begin{equation}
\label{Kap2.51}
\begin{split}
K(\omega,k_x,k_y;z) = i \omega - v \mu_1 \left( k_x \sigma_1 + k_y \sigma_2 - i \partial_z \sigma_3 \right) - m(z) \mu_3 - h \sigma_3, 
\end{split}
\end{equation}
is decomposed in the following way. 
We introduce the polar coordinates with respect to $(\omega, v k_x, v k_y)$, $ \omega = \rho \cos \theta , v k_x = \rho \sin \theta \cos \phi ,  v k_y = \rho \sin \theta \sin \phi$. 
Then (\ref{Kap2.51}) is decomposed as 
\begin{equation}
\begin{split}
K(\rho,\theta,\phi;z) &= i \rho \cos \theta- v \mu_1 \left( \rho \sin \theta \cos \phi \sigma_1 + \rho \sin \theta \sin \phi \sigma_2 - i \partial_z \sigma_3 \right) - m(z) \mu_3 - h \sigma_3 \\ 
&= U^{\dag} (\phi) \left[ i \rho \cos \theta - \rho \sin \theta \mu_1 \sigma_1 + i v \partial_z \mu_1 \sigma_3 - m(z) \mu_3 - h \sigma_3 \right] U(\phi) \\
&= U^{\dag} (\phi) \sigma_2 \left[ i \rho \cos \theta \sigma_2 + i \rho \sin \theta \mu_1 \sigma_3 - v \partial_z \mu_1 \sigma_1 - m(z) \mu_3 \sigma_2 - i h \sigma_1 \right] U(\phi) \\
&= U^{\dag} (\phi) \sigma_2 V^{\dag} (\theta) \left[ i \rho \sigma_2 - v \partial_z \mu_1 \sigma_1 - m(z) \mu_3 \sigma_2 - i h \sigma_1 \right] V(\theta) U(\phi) \\
&= U^{\dag} (\phi) \sigma_2 V^{\dag} (\theta) \sigma_2 \left[ i \rho + i v \partial_z \mu_1 \sigma_3 - m(z) \mu_3 - h \sigma_3 \right] V(\theta) U(\phi) \\
&= U^{\dag} (\phi) \sigma_2 V^{\dag} (\theta) \sigma_2 W^{\dag} \left[ i \rho + i v \partial_z \mu_1 \sigma_3 - m(z) \mu_2 - h \sigma_3 \right] W V(\theta) U(\phi), 
\end{split}
\end{equation}
where 
\begin{equation}
\begin{split}
U(\phi) = e^{\frac{i}{2}  \sigma_3\phi} ,
V(\theta) = e^{\frac{i}{2} \mu_1 \sigma_1 \theta} ,
W = \frac{1}{\sqrt{2}} (1 + i \mu_1) .
\end{split}
\end{equation}
Green's function is also decomposed as 
\begin{equation}
\begin{split}
&G(\rho,\theta,\phi;z_1,z_2) = U^{\dag}(\phi) V^{\dag}(\theta) W^{\dag} \tilde G(\rho,z_1,z_2) W \sigma_2 V(\theta) \sigma_2 U(\phi) .
\end{split}
\end{equation}
$\tilde G(\rho,z_1,z_2)$ is the Green's function for the one-dimensional Hamiltonian $\tilde H(z)$, 
\begin{equation}
\begin{split}
\tilde H(z) = - i v \partial_z \mu_1 \sigma_3 + m(z) \mu_2 + h \sigma_3. 
\end{split}
\end{equation}
$\tilde H(z)$ is diagonal in the spin space, 
$\tilde H (z) = \tilde H^{\uparrow }(z) \oplus \tilde H^{\downarrow }(z)$. 
Each of sectors forms the Jackiw-Rebbi model with the constant energy shift $\pm h$.

\section{An Analytically Solvable Model for the Jackiw-Rebbi problem}
\label{app3}
The Jackiw-Rebbi model with kink mass $m(z) = m \tanh (z/\xi)$ 
\begin{equation}
\begin{split}
H &= - i v \partial_z \mu_1 + m \tanh \left( z/\xi \right) \mu_2 \\
&=
\begin{pmatrix}
0 & -i \left( v \partial_z + m \tanh \left( z/\xi \right) \right) \\
-i \left( v \partial_z - m \tanh \left( z/\xi \right) \right) & 0
\end{pmatrix}
\end{split}
\label{JR}
\end{equation}
is analytically solvable.
In this appendix we briefly sketch how to solve Hamiltonian (\ref{JR}) according to ref.[\onlinecite{SK,TLM}]. 

The eigenvalue equations for $\Phi(z) = {}^t \left( u_+(z), u_-(z) \right)$ are 
\begin{equation}
\label{eq_for_u1}
\begin{cases}
-i \left( v \partial_z + m \tanh \left( z/\xi \right) \right) u_-(z) = \epsilon u_+(z) , \\
-i \left( v \partial_z - m \tanh \left( z/\xi \right) \right) u_+(z) = \epsilon u_-(z) .
\end{cases}
\end{equation}
For $\epsilon \neq 0$ we obtain 
\begin{equation}
\begin{split}
\left[ v^2 \partial^2_z + \epsilon^2 - m^2 \tanh^2(z/\xi) \mp \frac{m v}{\xi}\left( 1-\tanh^2(z/\xi) \right) \right] u_{\pm}(z) = 0 .
\end{split}
\label{eq_for_u2}
\end{equation}
By introducing a new variable $X=\tanh(z/\xi)$ , eq.(\ref{eq_for_u2}) may be rewritten as
\begin{equation}
\begin{split}
\left[ \left( 1-X^2 \right) \frac{d^2}{d X^2} - 2 X \frac{d}{d X} + \nu (\nu \mp 1) - \frac{\mu^2}{1-X^2} \right] u_{\pm} (X) = 0 , 
\end{split}
\end{equation}
where $\nu = \frac{m \xi}{v}, \mu^2 = \frac{\xi2}{v^2} (m^2-\epsilon^2)$.
The solutions of these differential equations are represented by Legendre functions $P^{\mu}_{\nu}(X)$, 
\begin{equation}
\begin{split}
&u_{+}(X) \sim P^{\mu}_{-\nu} (X) , \\
&u_{-}(X) \sim P^{\mu}_{\nu} (X) , 
\end{split}
\end{equation}
\begin{equation}
\begin{split}
P^{\mu}_{\nu}(X) = \frac{1}{\Gamma (1- \mu)} \left( \frac{1+X}{1-X} \right)^{\frac{\mu}{2}} F \left[ -\nu, \nu+1, 1-\mu ; \frac{1-X}{2} \right] , 
\end{split}
\end{equation}
where $F$ is the hypergeometric function, 
\begin{equation}
\begin{split}
& F[\alpha,\beta,\gamma,z] = \sum_{n=0}^{\infty} \frac{(\alpha)_n (\beta)_n}{(\gamma)_n} \frac{z^n}{n!}, \\
& (\alpha)_n = \alpha (\alpha+1) (\alpha+2) \cdots (\alpha + n -1).
\end{split}
\end{equation}
The relative phase factor between $u_{+}(X)$ and $u_{-}(X)$ is determined by eq.(\ref{eq_for_u1}). 
The eigenstates of which $\mu$ is pure imaginary correspond to continuum states, 
and can be labeled by the wave number $p$, and a sign of the energy $\pm$ :
\begin{equation}
\begin{split}
&\epsilon_{p,\pm} = \pm \sqrt{m^2 + v^2 p^2}, \\
&\Phi_{p,\pm}(z) = \frac{1}{\sqrt{2}}
\begin{pmatrix}
F\left[ \nu, 1-\nu, 1-i \xi p ; \frac{1-\tanh(z/\xi)}{2} \right] \\
\pm i \frac{m-i v p}{\sqrt{m^2 + v^2 p^2}} F\left[ -\nu, 1+\nu, 1-i \xi p ; \frac{1-\tanh(z/\xi)}{2} \right]
\end{pmatrix}
e^{i p z} .
\end{split}
\end{equation} 
Here, the normalization of $\Phi_{p,\pm}(z)$ is determined by the asymptotic behavior at $z \rightarrow \infty$, 
\begin{equation}
\begin{split}
\Phi_{p,\pm}(z) \rightarrow \frac{1}{\sqrt{2}}
\begin{pmatrix}
1 \\
\pm i \frac{m-i v p}{\sqrt{m^2 + v^2 p^2}}
\end{pmatrix}
e^{i p z} , \ \ (z \rightarrow \infty) .
\end{split}
\end{equation}
The bound states are determined by the expression of $P^{\mu}_{\nu}(X)$ as follows, 
\begin{equation}
\begin{split}
P^{\mu}_{\nu}(X) = \frac{2^{\mu}}{\Gamma (1- \mu)} \left( 1-X^2 \right)^{-\frac{\mu}{2}} F \left[ 1-\mu+\nu, -\mu-\nu, 1-\mu ; \frac{1-X}{2} \right] .
\end{split}
\end{equation}
This follows from the formula, 
\begin{equation}
\begin{split}
F\left[ \alpha, \beta, \gamma; z \right] = (1-z)^{\gamma-\alpha-\beta} F\left[ \gamma - \alpha, \gamma- \beta, \gamma; z \right] .
\end{split}
\end{equation}
To ensure that $F$ is finite
 for $X \rightarrow  -1$ (i.e. for $z \rightarrow -\infty$), 
we impose $\mu < 0 $.
$F$ must be a polynominal of a certain degree $n$, say, $-\mu - \nu = -n$. 
Then the non zero energy bound states can be labeled by integer 
$n = 1,2,\cdots,<\nu$ :
\begin{equation}
\begin{split}
&\epsilon_{n,\pm} = \pm m \sqrt{\frac{n}{\nu} \left( 2-\frac{n}{\nu} \right)} ,\\
&\Phi_{n,\pm}(z) = 
\begin{pmatrix}
\frac{\mp i}{\sqrt{2 \nu/n-1}} (1-X^2)^\frac{\nu-n}{2} F\left[ 1-n,2 \nu -n,1+\nu-n;\frac{1-X}{2} \right] \\
(1-X^2)^\frac{\nu-n}{2} F\left[ -n,1+2 \nu -n,1+\nu-n;\frac{1-X}{2} \right], 
\end{pmatrix} \\
&\hspace{33pt} = 
\begin{pmatrix}
\frac{\mp i}{\sqrt{2 \nu/n-1}} (1-X^2)^\frac{\nu-n}{2} \sum_{k=0}^{n-1} \frac{(1-n)_k (2 \nu-n)_k}{(1+\nu-n)_k} \frac{1}{k!} \left( \frac{1-X}{2}\right)^{k} \\
(1-X^2)^\frac{\nu-n}{2} \sum_{k=0}^{n} \frac{(-n)_k (1+2 \nu-n)_k}{(1+\nu-n)_k} \frac{1}{k!} \left( \frac{1-X}{2}\right)^{k} 
\end{pmatrix} , 
\end{split}
\end{equation}
up to normalization.
The zero energy bound state 
$\Phi_0 (z) \propto e^{-\int^z d z' m(z')/v } \sim 1/\left[ \cosh(z/\xi) \right]^{\nu} $ is 
\begin{equation}
\begin{split}
&\epsilon_0 = 0 , \\
&\Phi_0 (z) = 
\begin{pmatrix}
0 \\
1
\end{pmatrix}
\sqrt{\frac{\Gamma \left( \nu+1/2 \right)}{\xi \pi^{1/2} \Gamma \left( \nu \right) }} \ \frac{1}{\left[ \cosh \left( z/\xi \right) \right]^{\nu}} .
\end{split}
\end{equation}

In the case of $\nu = 1$ the expressions of eigenstates are considerably simplified. 
The continuum states are 
\begin{equation}
\begin{split}
&\epsilon_{p,\pm} = \pm \sqrt{m^2 + v^2 p^2} , \\
&\Phi_{p,\pm}(z) = \frac{1}{\sqrt{2}}
\begin{pmatrix}
1 \\
\pm \frac{v p + i m \tanh (\frac{m}{v} z)}{\sqrt{v^2 p^2 + m^2}}
\end{pmatrix}
e^{i p z}, 
\end{split}
\end{equation} 
and the zero energy state is the only allowed bound state, 
\begin{equation}
\begin{split}
&\epsilon_0 = 0 , \\
&\Phi_0 (z) = 
\begin{pmatrix}
0 \\
1
\end{pmatrix}
\sqrt{\frac{m}{2 v}} \ \frac{1}{\cosh \left(\frac{m z}{v} \right)}.
\end{split}
\end{equation}

\section{The Wigner Transformation of Green's Function $\tilde G (\rho ,z_1,z_2)$}
\label{app4}

In this appendix we perform the Winger transformation of $\tilde G (\rho ,z_1,z_2)$. 
For simplicity we set $v=1$. We calculate for the $\sigma_3 = \ \uparrow $ sector.
The calculation for the $\sigma_3 = \ \downarrow$ sector can be done in a similar way. The Green's function (\ref{G(z)}) is 
\begin{equation}
\begin{split}
& \tilde G^{\uparrow }(i \rho,z_1,z_2) \\
& \ \ \ = \int \frac{d p'}{2 \pi} 
\begin{pmatrix}
i\rho - h & p' - i m \tanh (m z_2) \\
p' + i m \tanh (m z_1) & (i\rho - h) \frac{(p' + i m \tanh (m z_1))(p' - i m \tanh (m z_2))}{p'^2 + m^2}
\end{pmatrix}
\frac{e^{ip'(z_1-z_2)}}{(i \rho - h)^2 - p'^2 - m^2} \\
& \ \ \ \ \ \ \ \  + 
\begin{pmatrix}
0 & 0 \\
0 & \frac{1}{i \rho - h}
\end{pmatrix}
\frac{m}{2} \mathrm{sech} (m z_1) \mathrm{sech} (m z_2) .
\end{split}
\end{equation}
The Wigner transformation of $\tilde G^{\uparrow }$ is defined by 
\begin{equation}
\begin{split}
G^{\uparrow } (i \rho, p,R) := \int d r \ \tilde G^{\uparrow }(i \rho,R+\frac{r}{2},R-\frac{r}{2}) e^{-i p r}. 
\end{split}
\end{equation}
In the integral of $r$, the nontrivial contributions stem from $r$-depending terms in 
$\tilde G^{\uparrow }(i \rho,R+\frac{r}{2},R-\frac{r}{2})$. There are four terms as shown below:
\begin{equation}
\begin{split}
(A) \ \ &\int dr \int \frac{d p'}{2 \pi}\  \frac{i m \tanh \left[ m (R+r/2) \right]}{(i\rho-h)^2-p'^2-m^2} e^{i p' r} e^{-i p r} , \\
(B) \ \ &\int dr \int \frac{d p'}{2 \pi}\  \frac{- i m \tanh \left[ m (R-r/2) \right]}{(i\rho-h)^2-p'^2-m^2} e^{i p' r} e^{-i p r} , \\
(C) \ \ &\int dr \int \frac{d p'}{2 \pi}\ \frac{( i \rho-h ) e^{i p' r} e^{-i p r}}{\left[ (i\rho-h)^2-p'^2-m^2 \right] (p'^2+m^2)} , \\
& \ \ \ \ \ \ \ \cdot \left[ i m p' \left( \tanh \left[  m (R+r/2) \right]- \tanh \left[  m (R-r/2) \right] \right) + m^2 \tanh \left[  m (R+r/2) \right] \tanh \left[  m (R-r/2) \right] \right]  \\
(D) \ \ &\int dr \ \frac{m}{2 (i \rho - h) } \mathrm{sech} \left[ m (R+r/2) \right] \mathrm{sech} \left[ m (R-r/2) \right] e^{-i p r} .
\end{split}
\label{appeq:abcd}
\end{equation}
The $p'$-integral can be performed by using the following formulae, 4
\begin{equation}
\begin{split}
&\int \frac{d p'}{2 \pi} \frac{e^{i p' r}}{p'^2 + A^2} = \frac{e^{-A |r|}}{2 A} ,\\
&\int \frac{d p'}{2 \pi} \frac{p' e^{i p' r}}{(p'^2 + A^2)(p'^2+m^2)} = - \frac{i}{2} \mathrm{sgn}(r)\  \frac{e^{-m|r|} - e^{-A |r|}}{(i \rho-h)^2} ,\\
&\int \frac{d p'}{2 \pi} \frac{e^{i p' r}}{(p'^2 + A^2)(p'^2+m^2)} = - \frac{1}{2 (i \rho - h)^2} \left( \frac{e^{-m|r|}}{m} - \frac{e^{-A |r|}}{A} \right) , 
\end{split}
\end{equation}  
where $A = \sqrt{m^2+(\rho + i h)^2}$. 
Then, the four terms in eq. (\ref{appeq:abcd}) are rewritten into the following forms,
\begin{equation}
\begin{split}
(A) &= - \frac{i m}{2 A} \int dr \ \tanh \left[ m (R+r/2) \right] e^{- A |r|} e^{-i p r}, \\
(B) &= \frac{i m}{2 A} \int dr \ \tanh \left[ m (R-r/2) \right] e^{- A |r|} e^{-i p r}, \\
(C) &= \frac{m}{2 (i \rho - h)} \int dr \ \mathrm{sgn} (r) \left\{ \tanh \left[m(R+r/2) \right] - \tanh \left[ m (R-r/2) \right] \right\} \left( e^{-m|r|} - e^{-A |r|} \right) e^{-i p r} \\
&\ \ \ \ \ - \frac{m^2}{2 (i \rho - h)} \int dr \ \left\{ \frac{ \tanh \left[  m (R+r/2) \right] + \tanh \left[ m (R-r/2) \right]}{\tanh (2 m R) } -1 \right\} \left( \frac{e^{-m|r|}}{m} - \frac{e^{-A |r|}}{A} \right) e^{-i p r}, \\
(D) &= \frac{m}{(i \rho - h) } \int dr \  \frac{e^{-i p r} }{ \cosh ( 2 m R ) + \cosh( m r )}. 
\end{split}
\end{equation}
We use the relations $\tanh(x) \tanh(y) = \frac{\tanh(x) + \tanh(y)}{\tanh(x+y)} - 1 $, and $\cosh(x) \cosh(y) = \left[\cosh(x+y) + \cosh(x-y)\right]/2$. 
After performing the Fourier transformation with respect to $r$, we obtain, 
\begin{equation}
\begin{split}
(A) &= - \frac{i}{2 A} \left[ f\left( m R, \frac{A+ip}{m}\right) - f\left( -m R, \frac{A-ip}{m}\right)  \right] ,\\
(B) &= \frac{i}{2 A} \left[ f\left( m R, \frac{A-ip}{m}\right) - f\left( m R, \frac{A-ip}{m}\right)  \right] ,\\
(C) &= \frac{1}{2 (i \rho - h)} 
\left[ f\left( m R, \frac{m+ip}{m}\right) + f\left( -m R, \frac{m-ip}{m}\right) + f\left( m R, \frac{m-ip}{m}\right) + f\left( -m R, \frac{m-ip}{m}\right) \right. \\
&\hspace{40pt} \left. - f\left( m R, \frac{A+ip}{m}\right) - f\left( -m R, \frac{A-ip}{m}\right) - f\left( m R, \frac{A-ip}{m}\right) - f\left( -m R, \frac{A-ip}{m}\right) \right] \\
& - \frac{1}{2 (i \rho - h) \tanh(2 m R)} 
\left[ f\left( m R, \frac{m+ip}{m}\right) + f\left( -m R, \frac{m-ip}{m}\right) - f\left( m R, \frac{m-ip}{m}\right) - f\left( -m R, \frac{m-ip}{m}\right) \right. \\
&\hspace{30pt} - \left.  \frac{m}{A} \left\{ f\left( m R, \frac{A+ip}{m}\right) + f\left( -m R, \frac{A-ip}{m}\right) - f\left( m R, \frac{A-ip}{m}\right) - f\left( -m R, \frac{A-ip}{m}\right) \right\} \right] \\
& - \frac{m^2 (i \rho - h)}{\left[ (i \rho - h)^2-p^2-m^2 \right] (p^2 + m^2)},  \\
(D) &= \frac{2 \pi \sin(2 p R)}{(i \rho - h) \sinh(2 m R) \sin \left(\pi p/m \right)} .
\end{split}
\end{equation}
Here $f(a,b)$ is defined by   
\begin{equation}
\begin{split}
f(a,b) := \int_0^{\infty} ds \ \tanh \left( a + \frac{s}{2} \right) e^{- b s } = \frac{1}{b} \left( -1 + 2 F[1,b,b+1,-e^{-2 a}]  \right) \ , \ \ \ (\Re (b) > 0) , 
\end{split}
\end{equation}
where $F[\alpha,\beta,\gamma,z]$ is the hypergeometric function. 
$F[1,b,b+1,z]$ is the analytic continuation of the infinite series, 
\begin{equation}
\begin{split}
F[1,b,b+1,z] = b \sum_{n=0}^{\infty} \frac{z^n}{b+n} \ , \ \ \ (|z|<1).
\end{split}
\end{equation}
We set an effective mass $\tilde m (\rho,p,R;h)$ and an effective factor $\beta (\rho,p,R;h)$ 
as follows, 
\begin{equation}
\begin{split}
&\tilde m (\rho,p,R;h) = \frac{p^2+A^2}{2 A} \left[f\left( m R, \frac{A+ip}{m}\right) - f\left( -m R, \frac{A-ip}{m}\right)\right] \\
&\beta (\rho,p,R;h) = \frac{p^2+A^2}{2 (i \rho - h)^2} 
\left[ f\left( m R, \frac{m+ip}{m}\right) + f\left( -m R, \frac{m-ip}{m}\right) + f\left( m R, \frac{m-ip}{m}\right) + f\left( -m R, \frac{m-ip}{m}\right) \right. \\
&\hspace{40pt} \left. - f\left( m R, \frac{A+ip}{m}\right) - f\left( -m R, \frac{A-ip}{m}\right) - f\left( m R, \frac{A-ip}{m}\right) - f\left( -m R, \frac{A-ip}{m}\right) \right] \\
& \hspace{30pt} - \frac{p^2+A^2}{2 (i \rho - h)^2 \tanh(2 m R)} 
\left[ f\left( m R, \frac{m+ip}{m}\right) + f\left( -m R, \frac{m-ip}{m}\right) - f\left( m R, \frac{m-ip}{m}\right) - f\left( -m R, \frac{m-ip}{m}\right) \right. \\
&\hspace{30pt} - \left.  \frac{m}{A} \left\{ f\left( m R, \frac{A+ip}{m}\right) + f\left( -m R, \frac{A-ip}{m}\right) - f\left( m R, \frac{A-ip}{m}\right) - f\left( -m R, \frac{A-ip}{m}\right) \right\} \right] \\
&\hspace{30pt} - \frac{p^2-m^2}{p^2+m^2} 
+ \frac{(i \rho-h)^2-p^2-m^2}{(i \rho-h)^2} \cdot \frac{2 \pi \sin(2 p R)}{\sinh(2 m R) \sin \left(\pi p/m \right)} ,
\end{split}
\end{equation}
where $A = \sqrt{m^2+(\rho + i h)^2}$ which depends on $h$. 
The Green's function for $\sigma_3 = \ \uparrow $ sector is represented as 
\begin{equation}
\begin{split}
G^{\uparrow }(i\rho,p,R) = \frac{1}{(i \rho-h)^2-p^2-m^2}
\begin{pmatrix}
i \rho-h & p - i \tilde m (\rho,-p,R;h) \\
p + i \tilde m (\rho,p,R;h) & (i \rho - h) \beta (\rho,p,R;h)
\end{pmatrix}.
\end{split}
\end{equation}
The effective mass $\tilde m$ and effective factor $\beta$ describe the spatial inhomogeneity 
near the kink structure.
In a similar way the Green's function for $\sigma_3 = \ \downarrow $ sector is represented as
\begin{equation}
\begin{split}
G^{\downarrow }(i\rho,p,R) = \frac{1}{(i \rho+h)^2-p^2-m^2}
\begin{pmatrix}
(i \rho + h) \beta (\rho,p,R;-h) & -p - i \tilde m (\rho,p,R;-h) \\
-p + i \tilde m (\rho,-p,R;-h) & i \rho+h
\end{pmatrix}.
\end{split}
\end{equation}

\section{The Derivation of Eq.(\ref{A(R)})}
\label{app5}

According to the polar decomposition (\ref{pd_G}), the integral with respect to $\phi $, and $\theta$ 
can be performed. 
First, we set $G(\rho,\theta,\phi,p_z,R_z) = U^{\dag}(\phi) G'(\rho,\theta,p_z,R_z) U(\phi)$, 
$G'(\rho,\theta,p_z,R_z) = V^{\dag}(\theta) W^{\dag} \tilde G(\rho,p_z,R_z) W \sigma_2 V(\theta) \sigma_2$.
Then
\begin{equation}
\begin{split}
G^{-1}\partial_{\mu}G &= U^{\dag}(\phi) G'^{-1}\partial_{\mu} G'(\rho,\theta,p_z,R_z) U(\phi)  \ ,\ \ \ \ (\mu = \rho,\theta,p_z,R_z) \ ,\\
G^{-1}\partial_{\phi}G &= \frac{i}{2} U^{\dag}(\phi) \left[ \sigma_3 - G'^{-1} \sigma_3 G' \right] U(\phi)\ .
\end{split}
\end{equation}
The integrand do not depend on $\phi$, 
\begin{equation}
\begin{split}
A_z(R_z) &= -\frac{i}{48 \pi^2} \int d\rho d\phi d\theta dp_z \ \epsilon^{\mu \nu \rho \sigma} \ \mathrm{tr} \left[ G^{-1}\partial_{\phi} G  G^{-1} \partial_{\mu} G G^{-1} \partial_{\nu} G G^{-1} \partial_{\rho} G G^{-1} \partial_{\sigma} G  \right] \\
&= \frac{1}{48 \pi} \int d\rho d\theta dp_z \ \epsilon^{\mu \nu \rho \sigma } \ \mathrm{tr} \left[ \left\{ \sigma_3 - G'^{-1} \sigma_3 G' \right\}  G^{-1} \partial_{\mu} G' G'^{-1} \partial_{\nu} G' G'^{-1} \partial_{\rho} G' G'^{-1} \partial_{\sigma} G'  \right] \\
&= \frac{1}{48 \pi} \int d\rho d\theta dp_z \ \epsilon^{\mu \nu \rho \sigma } \ \mathrm{tr} \left[  \sigma_3 \left\{ G^{-1} \partial_{\mu} G' G'^{-1} \partial_{\nu} G' G'^{-1} \partial_{\rho} G' G'^{-1} \partial_{\sigma} G' \right. \right. \\
& \hspace{150pt} \left. \left. -\partial_{\mu} G' G'^{-1} \partial_{\nu} G' G'^{-1} \partial_{\rho} G' G'^{-1} \partial_{\sigma} G' G^{-1} \right\} \right] \ ,
\end{split}
\end{equation}
where $(\mu, \nu, \rho, \sigma)$ run over $(\rho, \theta, p_z, R_z)$, and $\epsilon$ is fully 
anti-symmetric tensor with $\epsilon^{\rho \theta p_z R_z} = 1$.
Next, we calculate the the integral with respect to $\theta$. 
Due to the relations, $\sigma_2 V(\theta) \sigma_2 = \sigma_2 \left( \cos \frac{\theta}{2} + i \mu_1 \sigma_1 \sin \frac{\theta}{2} \right) \sigma_2 = \cos \frac{\theta}{2} - i \mu_1 \sigma_1 \sin \frac{\theta}{2} = V^{\dag}(\theta)$, we obtain,
\begin{equation}
\begin{split}
G'^{-1} \partial_{\mu} G' &= V(\theta) W^{\dag} \tilde G^{-1} \partial_{\mu} \tilde G W V^{\dag} (\theta),  \ \ \ \ (\mu \in (\rho,p_z,R_z)) \ ,\\
G'^{-1} \partial_{\theta} G' &= -\frac{i}{2} V(\theta) W^{\dag} \left[ \mu_1 \sigma_1 + \tilde G^{-1} \mu_1 \sigma_1 \tilde G \right] W V^{\dag}(\theta) \ , \\
\partial_{\mu} G'G'^{-1} &=V^{\dag} (\theta) W \partial_{\mu} \tilde G G^{-1}W^{\dag} V(\theta),\ \ \ \ (\mu \in (\rho,p_z,R_z)) \ , \\
\partial_{\theta} G'G'^{-1}&= -\frac{i}{2}V^{\dag}(\theta) W \left[ \mu_1 \sigma_1 + \tilde G \mu_1 \sigma_1 \tilde G ^{-1}\right] W^{\dag}  V(\theta)\ .
\end{split}
\end{equation}
Here, we introduce $\chi_{\mu},\xi_{\mu}$ to simplify notations as follows,
\begin{equation}
\begin{split}
\chi_{\mu} (\rho,p_z,R_z) &=  \tilde G^{-1} \partial_{\mu} \tilde G, \ \ \ \ (\mu \in (\rho,p_z,R_z)) \ , \\
\chi_{\theta}(\rho,p_z,R_z) &= \mu_1 \sigma_1 + \tilde G^{-1} \mu_1 \sigma_1 \tilde G \ , \\
\xi_{\mu}(\rho,p_z,R_z)&= \partial_{\mu} \tilde G G^{-1}, \ \ \ \ (\mu \in (\rho,p_z,R_z)) \ , \\
\xi_{\theta}(\rho,p_z,R_z) &= \mu_1 \sigma_1 + \tilde G \mu_1 \sigma_1 \tilde G ^{-1}\ .
\end{split}
\end{equation}
Then, we have,
\begin{equation}
\begin{split}
A_z(R_z) = -\frac{i}{96 \pi} \int d\rho d\theta dp_z \ \epsilon^{\mu \nu \rho \sigma } \ \mathrm{tr} \left[ \sigma_3 \left\{ V(\theta) \chi_{\mu}\chi_{\nu}\chi_{\rho}\chi_{\sigma} V^{\dag}(\theta) - V^{\dag}(\theta) \xi_{\mu}\xi_{\nu}\xi_{\rho}\xi_{\sigma} V(\theta) \right\} \right]\ .
\end{split}
\label{eq:F5}
\end{equation}
For $ V^{\dag}(\theta) \sigma_3  V(\theta) = \left( \cos \frac{\theta}{2} - i \mu_1 \sigma_1 \sin \frac{\theta}{2} \right) \sigma_3 \left( \cos \frac{\theta}{2} + i \mu_1 \sigma_1 \sin \frac{\theta}{2} \right) = \sigma_3 \left( \cos \theta + i \mu_1 \sigma_1 \sin \theta \right)$, 
and $ V(\theta) \sigma_3  V^{\dag}(\theta) = \left( \cos \frac{\theta}{2} + i \mu_1 \sigma_1 \sin \frac{\theta}{2} \right) \sigma_3 \left( \cos \frac{\theta}{2} - i \mu_1 \sigma_1 \sin \frac{\theta}{2} \right) = \sigma_3 \left( \cos \theta - i \mu_1 \sigma_1 \sin \theta \right)$, eq. (\ref{eq:F5}) is rewritten into,
\begin{equation}
\begin{split}
A_z(R_z) &= -\frac{i}{96 \pi} \int d\rho dp_z \int_{0}^{\pi} d\theta \ \epsilon^{\mu \nu \rho \sigma } \ \mathrm{tr} \left[ \sigma_3 \left\{ \left( \cos \theta + i \mu_1 \sigma_1 \sin \theta \right) \chi_{\mu}\chi_{\nu}\chi_{\rho}\chi_{\sigma} - \left( \cos \theta - i \mu_1 \sigma_1 \sin \theta \right) \xi_{\mu}\xi_{\nu}\xi_{\rho}\xi_{\sigma} \right\} \right] \\
&= -\frac{i}{96 \pi} \int d\rho dp_z \ \epsilon^{\mu \nu \rho \sigma } \ \mathrm{tr} \left[ \sigma_3 \left\{ 2 i \mu_1 \sigma_1 \chi_{\mu}\chi_{\nu}\chi_{\rho}\chi_{\sigma} + 2 i \mu_1 \sigma_1 \xi_{\mu}\xi_{\nu}\xi_{\rho}\xi_{\sigma} \right\} \right] \\
&= \frac{i}{48 \pi} \int d\rho dp_z \ \epsilon^{\mu \nu \rho \sigma } \ \mathrm{tr} \left[ \mu_1 \sigma_2 \left\{ \chi_{\mu}\chi_{\nu}\chi_{\rho}\chi_{\sigma} + \xi_{\mu}\xi_{\nu}\xi_{\rho}\xi_{\sigma} \right\} \right] \ . 
\end{split}
\end{equation}

\end{document}